\documentclass[nofootinbib,twocolumn,preprintnumbers]{revtex4-1}
\usepackage[dvips]{graphicx}
\usepackage{amsmath,amsthm,amssymb,multirow,psfrag}
\begin{document}

\def\lsim{\mathrel{\rlap{\lower4pt\hbox{\hskip1pt$\sim$}}
    \raise1pt\hbox{$<$}}}
\def\gsim{\mathrel{\rlap{\lower4pt\hbox{\hskip1pt$\sim$}}
    \raise1pt\hbox{$>$}}}
\newcommand{\vev}[1]{ \left\langle {#1} \right\rangle }
\newcommand{\bra}[1]{ \langle {#1} | }
\newcommand{\ket}[1]{ | {#1} \rangle }
\newcommand{\ev}{ {\rm eV} }
\newcommand{\kev}{{\rm keV}}
\newcommand{\mev}{{\rm MeV}}
\newcommand{\gev}{{\rm GeV}}
\newcommand{\tev}{{\rm TeV}}
\newcommand{\mpl}{$M_{Pl}$}
\newcommand{\mw}{$M_{W}$}
\newcommand{\Ft}{F_{T}}
\newcommand{\Zparity}{\mathbb{Z}_2}
\newcommand{\BLambda}{\boldsymbol{\lambda}}
\newcommand{\be}{\begin{eqnarray}}
\newcommand{\ee}{\end{eqnarray}}
\newcommand{\met}{\;\not\!\!\!{E}_T}

\title{Directly Measuring the Tensor Structure of the Scalar Coupling to Gauge Bosons}
\author{Daniel Stolarski$^{1,2}$ and Roberto Vega-Morales$^{3,4}$}
\affiliation{$^1$Department of Physics and Astronomy, Johns Hopkins University, Baltimore, MD 21218\\  
$^2$Center for Fundamental Physics, Department of Physics, University of Maryland, College Park, MD 20742\\
$^3$ Department of Physics and Astronomy, Northwestern University, Evanston, IL 60208\\
$^4$ Theoretical Physics Department, Fermilab, Batavia, IL 60510, USA}

\begin{abstract}

Kinematic distributions in the decays of the newly discovered resonance to four leptons can provide a direct measurement of the tensor structure of the particle's couplings to gauge bosons. Even if the particle is shown to be a parity even scalar, measuring this tensor structure is a necessary step in determining if this particle is responsible for giving mass to the $Z$. We consider a Standard Model like coupling as well as coupling via a dimension five operator to either $ZZ$ or $Z\gamma$.  We show that using full kinematic information from each event allows discrimination between renormalizable and higher dimensional coupling to $ZZ$ at the 95\% confidence level with ${\mathcal O}(50)$ signal events, and coupling to $Z\gamma$ can be distinguished with as few as 20 signal events. This shows that these measurements can be useful even with this year's LHC data.  

\end{abstract}

\preprint{FERMILAB-PUB-12-499-PPD-T, NUHEP-TH/12-10, UMD-PP-012-018, CAS-KITPC/ITP-343}

\maketitle

\section{Introduction}

The discovery of a new particle by the CMS~\cite{CMSHiggs} and ATLAS~\cite{ATLASHiggs} experiments is a great triumph for particle physics, but it also leads to a host of new questions about the nature of the new state.  The question of whether this particle gives mass the the  $W$ and $Z$ is of paramount importance.   Since the discovery, the theoretical community has begun to weigh in on this question with many different analyses~\cite{Low:2012rj,Corbett:2012dm,Arbey:2012dq,Giardino:2012dp,Buckley:2012em,Baglio:2012et,Gunion:2012gc,Ellis:2012hz,Montull:2012ik,Espinosa:2012im,Carmi:2012in,Cohen:2012wg,Banerjee:2012xc,Cao:2012yn,Bertolini:2012gu,Joglekar:2012hb,ArkaniHamed:2012kq,Bonnet:2012nm,Craig:2012vn,Matsuzaki:2012mk,Moffat:2012pb,Plehn:2012ab,Dorsner:2012ab,Coleppa:2012eh}. These analyses show that the new state is broadly consistent with the Standard Model Higgs, but all of these analyses use only the information obtained from cross sections and branching ratios. While this form of analysis is very useful, it is ultimately a model dependent test of the properties of the new state. 

Here we attempt to delve into the signal events themselves and see what information can be learned about the new particle from the kinematic distributions of the final state particles. Specifically, the channel where the new particle decays to  four leptons via intermediate gauge bosons~\cite{CMS4lep,ATLAS4lep} contains a tremendous amount of information about the new resonance. It has been shown that this channel can be used to distinguish the parity and spin~\cite{Soni:1993jc,Barger:1993wt,Choi:2002jk,Buszello:2002uu,Keung:2008ve,Antipin:2008hj,Gao:2010qx,DeRujula:2010ys,DeSanctis:2011yc} of a new resonance, with~\cite{Gao:2010qx} and~\cite{DeRujula:2010ys} doing an exhaustive comparison of many different spin and parity possibilities.  It has also been shown that kinematic methods can be used to distinguish signal from background in this channel~\cite{Gainer:2011xz}. 

Here we take the hypothesis that the new particle is a parity even scalar and try to see if this channel can be used to directly measure the tensor structure of the coupling of this particle to the four lepton final state.  If we denote the new scalar by $\phi$, it can have the following couplings to $ZZ$
\begin{eqnarray}
 \frac{1}{v} \Big(&a_h&\, m_{Z}^2\, \phi\, Z_\mu Z^\mu + a_s\, \phi\, Z^{\mu\nu} Z_{\mu\nu} 
 + ... \Big)
\label{eq:zzlag}
\end{eqnarray}
where $Z_\mu$ is the $Z$ field while $Z_{\mu\nu} = \partial_\mu Z_\nu - \partial_\nu Z_\mu$.
  Here $v=246$ GeV is the Standard Model (SM) Higgs vev which is chosen to normalize the operators, and the $...$ is for operators of dimension higher than five. If $\phi$ is the Standard Model Higgs, then $a_h=i$, and the other coupling is loop induced and small.  

As we are trying to determine whether this new particle is the SM Higgs, we must consider other possibilities. If $\phi$ does not give mass to the $Z$, then its linear coupling to gauge bosons can proceed via the field strength tensor, $Z_{\mu\nu}$ as in the operator $a_s$ in Eq.~(\ref{eq:zzlag}). There are many such models in the literature, see for example~\cite{DeRujula:2010ys,Low:2010jp,Fox:2011qc} and references therein.  The $a_s$ operator is generically loop induced and its coefficient is model dependent. We see this sort of operator even in the Standard Model Higgs' coupling to $\gamma\gamma$ and $Z\gamma$:
\begin{eqnarray}
\frac{1}{v} \Big( a_\gamma\, \phi\, F^{\mu\nu} F_{\mu\nu} +  a_{Z\gamma} \, \phi\, Z^{\mu\nu} F_{\mu\nu}  
                  + ... \Big)
\label{eq:zgam}                  
\end{eqnarray}
where we continue to use $\phi$ to denote our scalar, and $F_{\mu\nu}$ is the field strength tensor for the photon. In the SM, $a_\gamma$ and $a_{Z\gamma}$ are induced by loops with top and $W$ giving the largest contributions. If $\phi$ is not the Higgs, then a plausible alternative is that it decays to four leptons via $a_s$ or $a_{Z\gamma}$. Generically, $a_s$, $a_{Z\gamma}$ and $a_\gamma$ are all present and of comparable size, and all three operators can mediate four lepton final states. The experimental searches~\cite{CMS4lep,ATLAS4lep} require that the invariant mass of the one of the lepton pairs is near the $Z$ pole, so the contribution of $a_\gamma$ is small, but we will see that both $a_s$ and $a_{Z\gamma}$ need to be considered. 

Bounds on all the operators in Eqs.~(\ref{eq:zzlag}) and~(\ref{eq:zgam}) can be set using the absence of single production of this resonance at LEP~\cite{Hankele:2006ma}. In order to interpret these bounds at the LHC, however, the production cross section of this scalar must be computed, and that is \textit{a priori} unknown. 

If $\phi$ does couple dominantly via $a_h$ that would be evidence that it is indeed a Higgs.  On the other hand, it could still be something more exotic such as a dilation~\cite{Gildener:1976ih,Goldberger:2007zk,Fan:2008jk} or a radion~\cite{Goldberger:1999uk}.  The crucial point is that if we are going determine if $\phi$ gives mass to the $Z$ boson, we must show that its coupling to $ZZ^*$ is dominantly through $a_h$.  As we will show in this paper, the kinematic distributions of the four lepton events can discriminate $a_h$ from $a_s$ and $a_{Z\gamma}$. 

The question of distinguishing these operators via kinematics was considered briefly in~\cite{Cao:2009ah}. We here extend their analysis by studying all possible kinematic variables which can distinguish different possible decay operators.  Furthermore, the analysis of~\cite{Cao:2009ah} only considers a mass of a scalar greater than twice the $Z$ mass, while we here will be working in the kinematic regime where one of the $Z$'s is far off-shell.  In~\cite{DeRujula:2010ys}, they use kinematic methods to distinguish two different kinds of parity even scalars, but both of their possibilities are still responsible for giving mass to the $W$ and $Z$. In other words, they both have significant $a_h$ in the language of Eq.~(\ref{eq:zzlag}).

The analysis of~\cite{Low:2012rj} shows that with a fit of the $\gamma\gamma$, $ZZ^*$, and $WW^*$ rates, as well as the absence of a large anomaly in continuum $Z\gamma$, that the scenario of the four lepton decays being due to $a_s$ is strongly disfavored. As already mentioned, while this statement has few assumptions, it is still model dependent and we seek to confirm this exclusion by a direct measurement.  

If the new resonance has anomalous couplings to $Z$ and $W$, then the production of the resonance through vector boson fusion (VBF) would also be modified. These effects were studied in~\cite{Plehn:2001nj,Hankele:2006ma,Klamke:2007cu,Hagiwara:2009wt}, where it was shown that angular correlations between the two tagging jets in VBF can constrain the value of the operators in Eq.~(\ref{eq:zzlag}).\footnote{Monte Carlo code which simulates anomalous VBF events can be found in~\cite{Arnold:2012xn}.} Here, we only consider decays of the resonance, but these two types of measurements can be complementary in fully characterizing the nature of the new state.

The organization of this paper is as follows: in Sec.~\ref{sec:events} we describe the kinematics of four lepton events, as well as our event generation and selection cuts. We also show the distributions which will allow discrimination of the different scenarios. In Sec.~\ref{sec:stats} we describe the statistical procedure we use to distinguish the different scenarios, and in Sec.~\ref{sec:conc} we conclude.

\section{Four Lepton Events}
\label{sec:events}

The kinematics of four lepton events are described in detail in many places in the literature, see for example~\cite{DeRujula:2010ys,Gainer:2011xz}. Here we describe only the variables relevant to our analysis.  Because we are trying to distinguish different scalar scenarios, information from the production vertex is lost, and angles relative to the beam are irrelevant.  The kinematic variables sensitive to decays of the scalar are:

\begin{itemize}

\item $\Phi$ -- The angle between the decay planes of the two $Z$ bosons in the rest frame of the scalar.
\item $\theta_1$ -- The angle between the fermion coming from the decay of $Z_1$ and the $Z_2$ momentum in the $Z_1$ rest frame. 
\item $\theta_2$ -- $\theta_1$ with $Z$'s interchanged.
\item $M_i$ -- The invariant mass of the two $Z$'s. We take the convention $M_1 > M_2$. 

\end{itemize}
The distributions for $\theta_1$ and $\theta_2$ are the same because of an exchange symmetry between the two $Z$'s when they both decay to leptons, but it is important to use both because that increases the number of observables for each event. In most events, $M_1 \sim M_Z$ regardless of the coupling to the $Z$.  These variables are all independent subject to the constraint $(M_1 + M_2) \leq \sqrt{s}$ where $s$ is the invariant mass squared of the four lepton system.  We consider four electron, four muon, and two electron + two muon events. In the first two types of events there is a two fold ambiguity in assigning leptons to parent $Z$'s, which we break by taking $M_1$ to be the pair of opposite sign leptons whose invariant mass is closest to the $Z$. 

\begin{figure}
\includegraphics[width=0.45\textwidth]{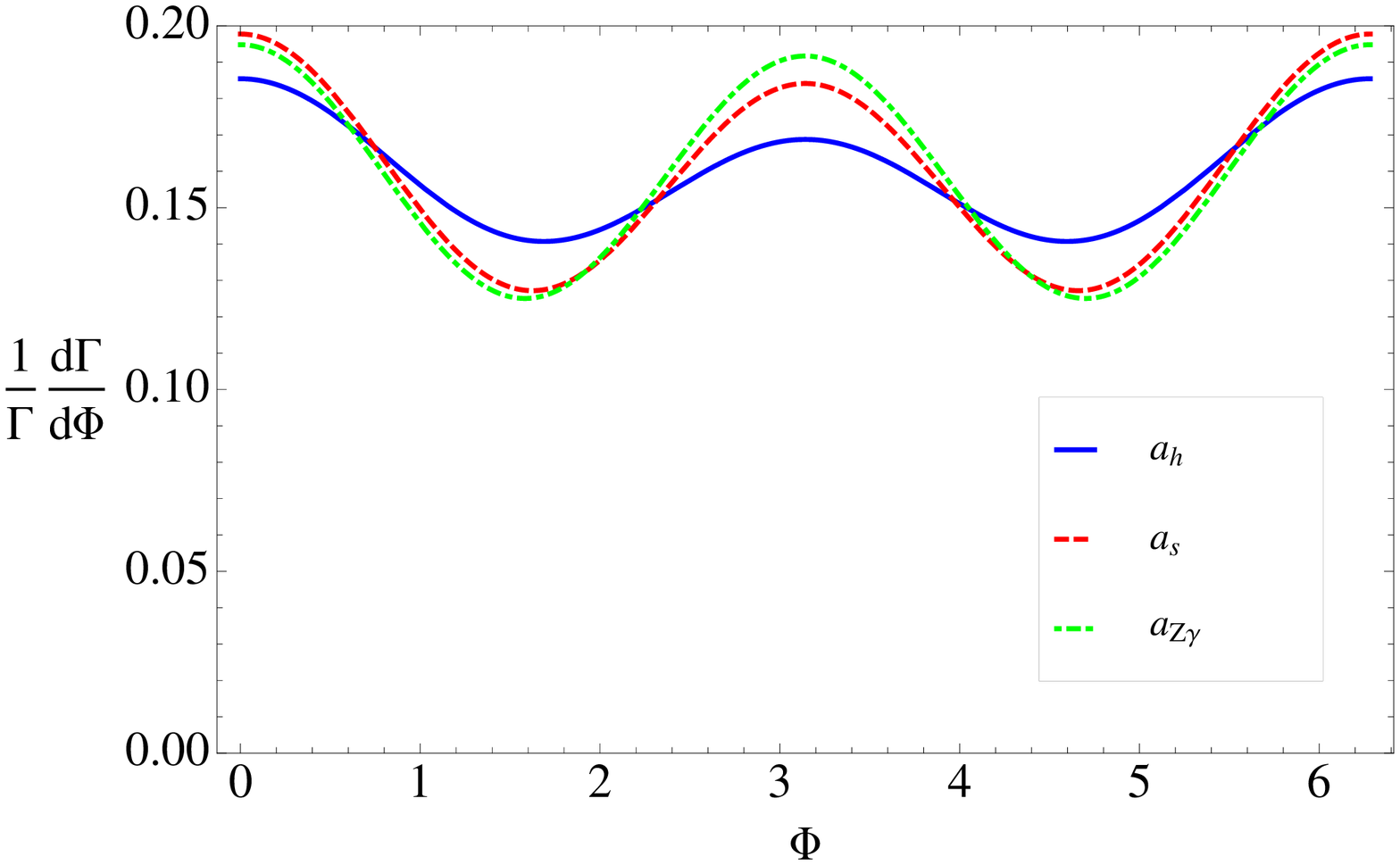}
\includegraphics[width=0.45\textwidth]{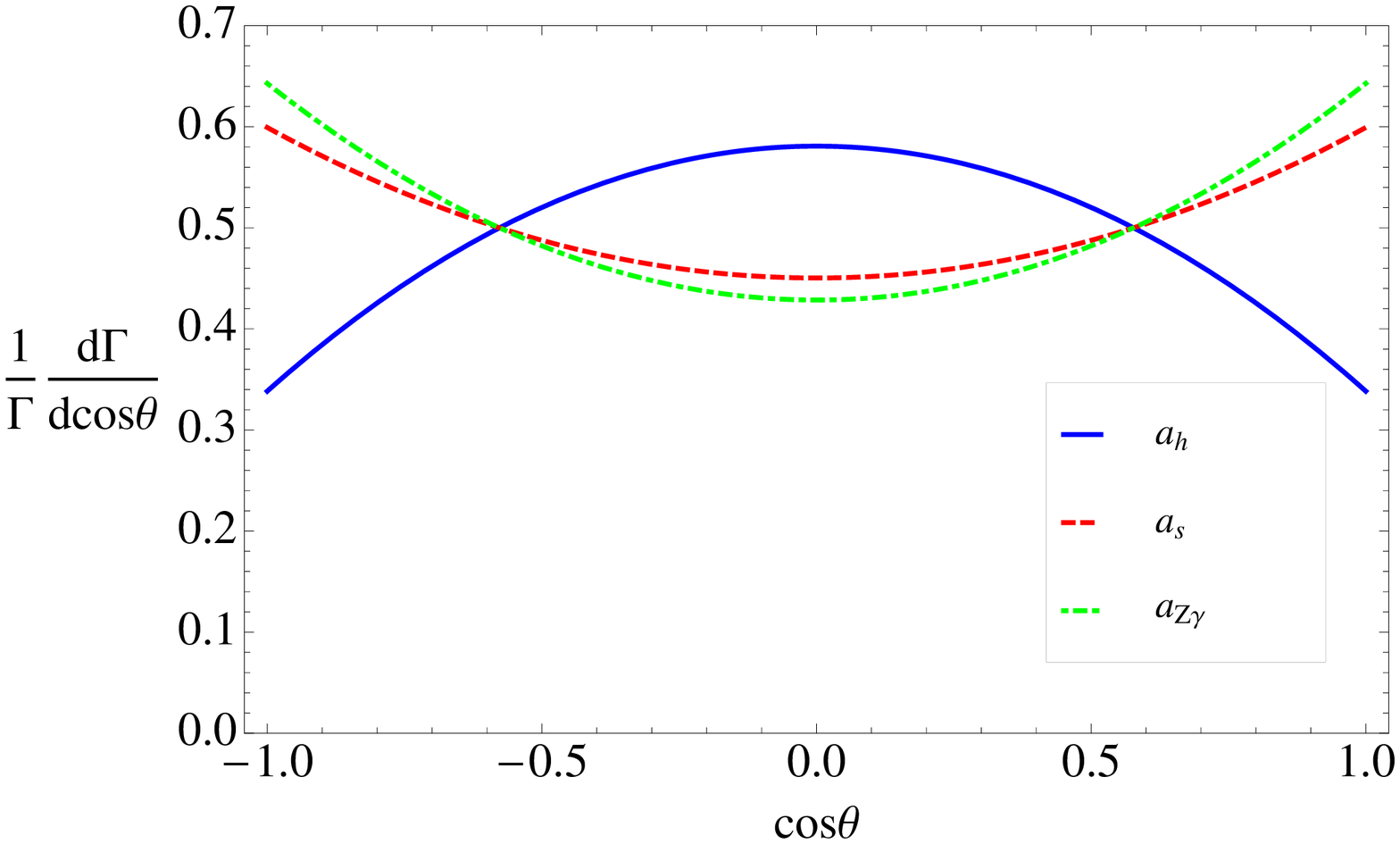}
\includegraphics[width=0.45\textwidth]{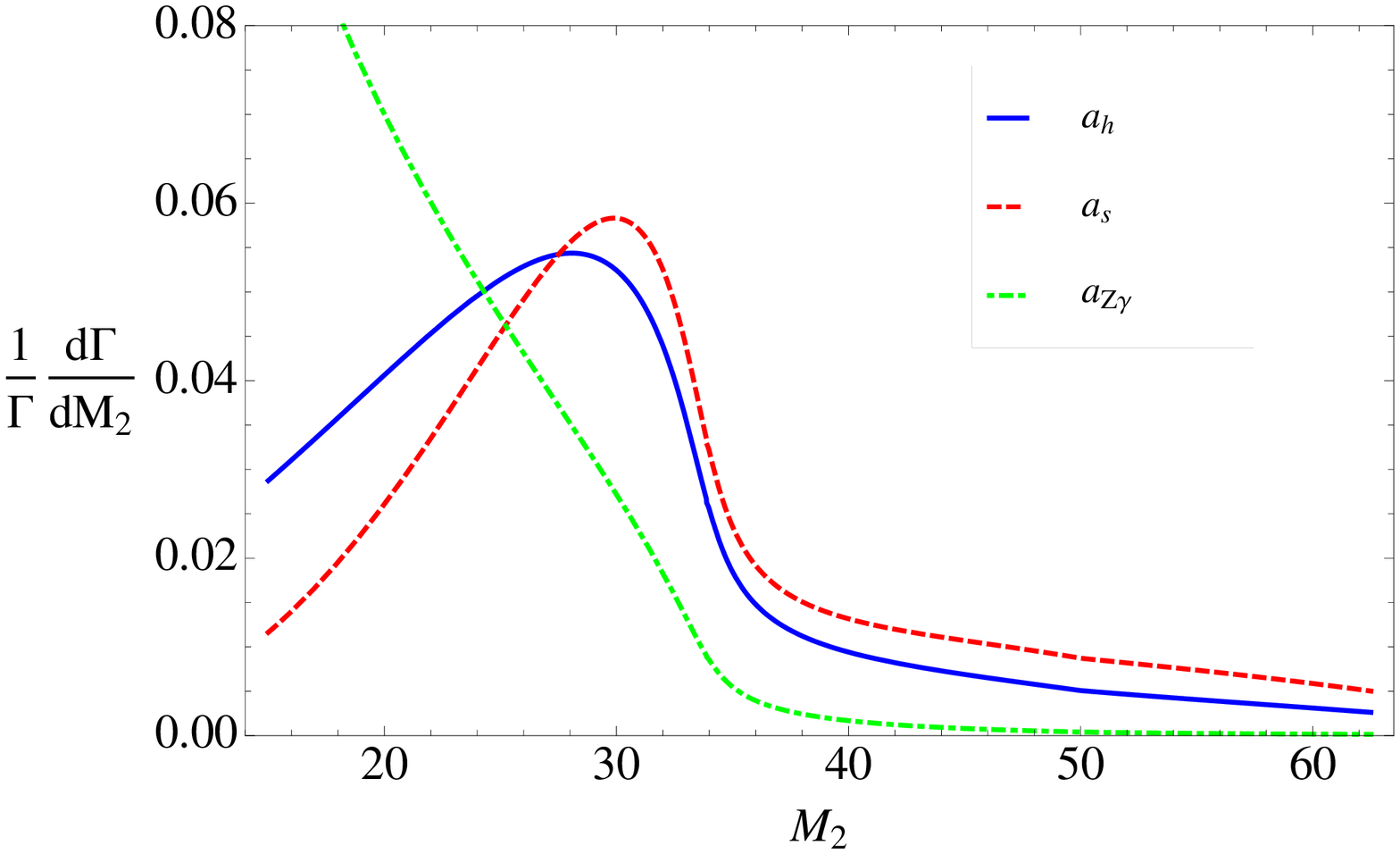}
\caption{Normalized distributions for $\Phi$ (top), $\cos\theta_i$ (middle), and $M_2$ (bottom) for $m_\phi = 125$ GeV.  Each plot shows curves from our three different scenarios with  $a_h$ blue (solid), $a_s$ red (dashed), and $a_{Z\gamma}$ green (dot-dashed). }
\label{fig:distributions}
\end{figure}

We compute tree level analytic expressions for the full differential decay width in terms of frame invariant 4-vector dot products before choosing the frame with kinematic variables described above.  These expressions, at least of the Higgs, can be found in many places in the literature including~\cite{Gainer:2011xz} and references therein.  NLO and finite mass corrections have been computed in the case of the Higgs~\cite{Kniehl:2012rz} and have been shown to be a few per cent.  We plot normalized one dimensional distributions for $\Phi$, $\cos\theta_i$, and $M_2$ in Fig.~\ref{fig:distributions}. We take $m_\phi = 125$ GeV here and throughout.  Distributions will change little with variations of $m_\phi$ within the experimental resolution. We restrict $M_i$ to the ranges described below in order to better approximate the experimental searches.  We consider the operators $a_h$, $a_s$ and $a_{Z\gamma}$ turning on one operator at a time with the others set to zero. 

In general, all three operators, $a_h$, $a_s$, and $a_{Z\gamma}$ will be non-zero.  In the case of a Higgs-like state which gives mass to the $Z$, $a_s$ and $a_{Z\gamma}$ are loop suppressed and $a_h$ will dominate. If the new state does not contribute to electroweak symmetry breaking, then $a_h$ will often be negligible.  If $a_s\sim a_{Z\gamma}$, which is typically the case if the two operators are generated by loops of electroweak charged matter, then the effects of $a_{Z\gamma}$ will dominate. This is because in the allowed region for $M_2$, a photon will be much closer to on-shell than a $Z$, and because the $Z$ has suppressed couplings to leptons relative to the photon.  Even if $a_s$ is ten times larger than $a_{Z\gamma}$, $a_{Z\gamma}$ will dominate the decay and we can consider turning on just $a_{Z\gamma}$ as a reasonable approximation.  On the other hand, one could imagine a model where $a_{Z\gamma}$ is very small, possibly due to tuning, and we therefore consider turning on only $a_s$ as a stand in for this possibility. From this analysis, we see that in most of the parameter space, one operator will dominate over the other two, which is why we consider scenarios where only one operator is turned on at a time. 

The top two panels in Fig.~\ref{fig:distributions} show that the angular distributions, particularly that of $\cos\theta$ provide good discriminating power between a Higgs-like scenario $a_h$, and the two non-Higgs-like possibilities.  The third plot shows that the $M_2$ distribution is different for all three scenarios, and the difference is even more pronounced for small values of $M_2$. This can be seen from the following simple analysis.  For $a_h$, the matrix element goes to a constant as $M_2 \rightarrow 0$, and a phase space factor of $M_2 dM_2$ makes the rate go to zero.  For $a_s$, the matrix element goes as $M_2$ for small $M_2$ because of the derivative in the operator, so $d\Gamma$ falls as $M_2^3$. Finally, for $a_{Z\gamma}$, the matrix element goes as $1/M_2$ because the photon propagator in the denominator and the derivative in the numerator, and thus the rate goes as $1/M_2$. As we will see below, realistic detector cuts such as those on lepton $p_T$ will change this low $M_2$ behavior, but this simple analysis shows that if the experiments could push down the $M_2$ reach of the events, they would gain discriminatory power. 

We do not include a plot for $M_1$ because in all scenarios, it looks similar with a large peak at $M_Z$ that has width of $\Gamma_Z$. The $M_1$ distribution does, however, provide some discrimination power in that the number of events well below $M_Z$ differs for our three different scenarios.  For example, in the $a_h$ scenario, 70\% of the events will lie more than $2\Gamma_Z$ away from $M_Z$, while the corresponding fraction for $a_s$ ($a_{Z\gamma}$) is 64\% (84\%).  	The majority of these non-resonant events have $M_1 < M_Z$. 

If the four lepton events are dominated by $a_{Z\gamma}$, then there should also be decays to on-shell photons.  It has been pointed out that searching for the Higgs in decays to $Z\gamma$ is a promising channel~\cite{Gainer:2011aa}. While there is as yet no direct limit in this channel, \cite{Low:2012rj} uses the measurement of the $Z\gamma$ cross section to place a limit on the ratio of the $Z\gamma$ mode to the four lepton mode to be about 40. Given this, we take the $Z\gamma$ mode to be an unlikely possibility, but we still believe in checking the data to see if it can be directly excluded. 

\begin{figure}
\includegraphics[width=0.45\textwidth]{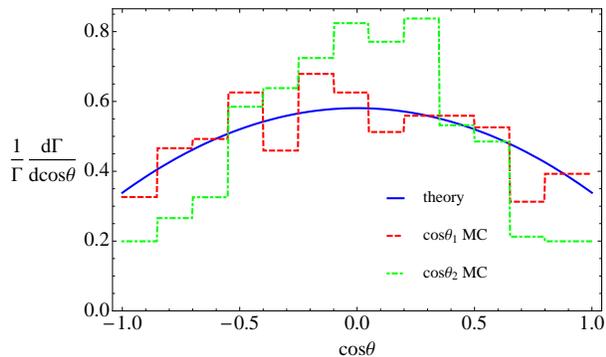}
\caption{Normalized distribution for $\cos\theta$ in the $a_h$ scenario. The blue (solid) curve is the same as the theory curve from Fig.~\ref{fig:distributions}, the red (dashed) histogram is the distribution for $\cos\theta_1$ for 1000 Monte Carlo events, while the green (dot-dashed) histogram is $\cos\theta_2$ for the same events.  }
\label{fig:costhetaMC}
\end{figure}

\begin{figure}
\includegraphics[width=0.45\textwidth]{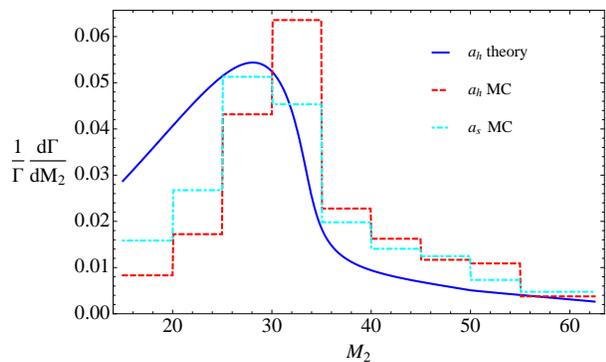}
\caption{Normalized $M_2$ distributions. The blue (solid) curve is the theory prediction in the $a_h$ scenario, while the light blue (dot-dashed) histogram is 1000 Monte Carlo events also in the $a_h$ scenario. The red (dashed) histogram is 1000 events in the $a_s$ scenario. }
\label{fig:M2MC}
\end{figure}

In order to compare to experiment, we also generate Monte Carlo (MC) events. We use the Johns Hopkins MC described in~\cite{Gao:2010qx} to simulate $a_h$ and $a_s$, and Madgraph 5~\cite{Alwall:2011uj} for $a_{Z\gamma}$. We generate $gg \rightarrow \phi \rightarrow 4 \ell$ events where $\ell = e,\mu$ at the LHC with $\sqrt{s} = 8$ TeV. Gluon fusion is the dominant mode of Higgs production at the LHC~\cite{Dittmaier:2011ti}. Since our variables are mostly sensitive to decay and not production, the errors introduced by ignoring sub-dominant production modes will be small. We require our events to contain four charged leptons ($e$ or $\mu$) with 
\begin{itemize}
\item $p_T > 10$ GeV
\item $|\eta| < 2.5$
\item$50 \, {\rm GeV} < M_1 < 110 \, {\rm GeV} $
\item $M_2 > 15$ GeV,
\end{itemize}
which roughly mimics the experimental selection criteria in~\cite{CMS4lep,ATLAS4lep}.   Histograms for the distinguishing kinematic variables from generated events are overlaid on the analytic results in Figs.~\ref{fig:costhetaMC} and~\ref{fig:M2MC}.  Because the experimental resolution for energy and direction of leptons is so precise, we do not apply any smearing to the events. While a truly realistic study will need to take into account experimental reality, we here see how far the experiments could get with just the geometric cuts above. 

In Fig.~\ref{fig:costhetaMC}, we plot the $\cos\theta_1$ and $\cos\theta_2$ distributions for 1000 generated Monte Carlo events which pass the above cuts. We compare it to the theoretical distribution which is the same for the two angles. We see that the cuts have limited effect on $\cos\theta_1$, but the rate for $\cos\theta_2 \sim \pm 1$ is suppressed. This is because in that configuration, one of the leptons is nearly aligned with the boost direction needed to go to the lab frame from the $Z_2$ rest frame, and thus preforming that boost will reduce its energy and make it less likely to pass the $p_T$ cut. This effect is small for $\cos\theta_1$ because the lepton energies in the $Z_1$ rest frame are much larger. 

In Fig.~\ref{fig:M2MC}, comparing the blue (solid) curve to the light-blue (dot-dashed) histogram, we see that the experimental cuts reduce the event rate for small $M_2$.  Even after these cuts, however, the histograms for $a_h$ and $a_s$ still differ, so the experimental cuts do not wash out the discriminating power.

\section{Distinguishing Operators}
\label{sec:stats}

In order to estimate the ability of the LHC to discriminate a Higgs-like scenario dominated by $a_h$ from other scenarios, we employ a likelihood analysis of the generated events. We consider only signal events because requiring the invariant mass of the four lepton system to be near the mass of the new boson can make the signal to background ratio significantly larger than one. Furthermore, reweighting techniques such as the one laid out in~\cite{Pivk:2004bb} can be used to further purify the event selection. 

We use a standard unbinned likelihood analysis which is described in detail in~\cite{Gao:2010qx}. We can use the computed normalized differential cross section as a probability distribution $P(\Phi,\theta_i,M_i|a_i)$ for each operator $a_h$, $a_s$, and $a_{Z\gamma}$. The normalization is computed with the $M_i$ cuts described above because they are independent of Lorentz frame.  Taking the $p_T$ and $\eta$ acceptance into account in $P$ would improve the statistical power of the test, but because those cuts are frame-dependent, we leave that to further work. 

Given a sample of $N$ events, we can then construct a likelihood ${\cal L}(a_i)= \prod_{j=1}^N P_j(a_i)$. With this likelihood we can then compare two different scenarios, $a_1$ and $a_2$ by constructing a hypothesis test with test statistic defined by~\cite{Cousins:2005pq}
\begin{equation}
\Lambda = 2\log[{\cal L}(a_1)/{\cal L}(a_2)].
\end{equation}
Since we are taking the resonance mass as input and using the normalized differential cross sections to construct our likelihood functions, there are no free parameters (nuisance parameters) in this ratio, making this a simple hypothesis test. 

\begin{figure}
\includegraphics[width=0.45\textwidth]{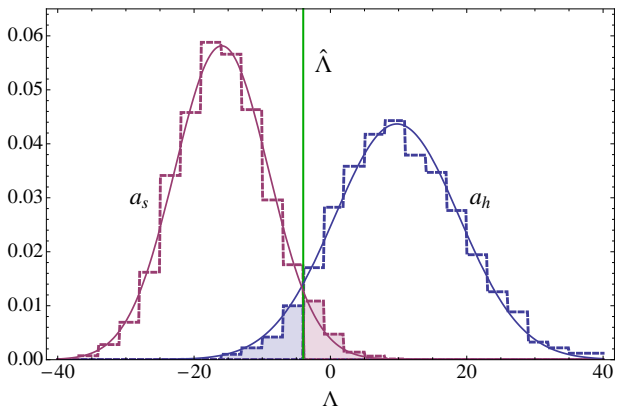}
\caption{Normalized distribution of our test statistic $\Lambda$ when $a_h$ is true on the right (blue), and when $a_s$ is true on the left (pink).  Each histogram is the result of 5000 pseudo-experiments with 50 events each. The vertical (green) line is $\hat \Lambda$ defined in Eq.~(\ref{eq:lambdahat}) such that the area to the right of $\hat \Lambda$ under the $a_s$ histogram is equal to the area to the left of $\hat \Lambda$ under the $a_h$ histogram. We also draw a Gaussian over each histogram with the same median and standard deviation.  }
\label{fig:lambda}
\end{figure}

To estimate the expected significance of discriminating between two different hypotheses corresponding to two different operators, we follow a similar analysis to that found in~\cite{Gao:2010qx}.  To begin, we take one hypothesis as true, say $a_1$ and generate a fixed number $N$ of $a_1$ events.  We then construct $\Lambda$ as above for a large number of pseudo-experiments each containing $N$ events in order to obtain a distribution for $\Lambda$.  We then repeat this exercise taking $a_2$ to be true and again obtain a distribution for $\Lambda$. These two distributions are shown in Fig.~\ref{fig:lambda} comparing $a_h$ and $a_s$. This figure shows 5000 pseudo-experiments of 50 events each, which shows a clear separation between the two scenarios.  

With the two distributions for $\Lambda$ in hand we can compute an approximate significance by the following procedure. If we denote the distribution with negative mean as $f$ and the distribution with positive mean as $g$, we find a value  $\hat \Lambda$ such that 
\begin{equation}
\int_{\hat \Lambda}^\infty fdx = \int_{-\infty}^{\hat \Lambda} gdx.
\label{eq:lambdahat}
\end{equation}
Schematically, this value of $\hat \Lambda$ corresponds to a value such that if the experiment observed that value for the test statistic, it would have no discriminatory power between the two scenarios.  We then interpret the probability given by either side of Eq.~(\ref{eq:lambdahat}) as a one sided Gaussian probability, which can then be interpreted in terms of number of $\sigma$. This procedure is shown schematically in Fig.~\ref{fig:lambda} with the areas of the two shaded regions being equal and corresponding to the probability of excluding the correct hypothesis.  For a simple hypothesis test, this Gaussian approximation is often sufficient~\cite{Cousins:2005pq}, and we see from Fig.~\ref{fig:lambda} that the $\Lambda$ distributions are well approximated by Gaussians. 

\begin{figure}
\includegraphics[width=0.45\textwidth]{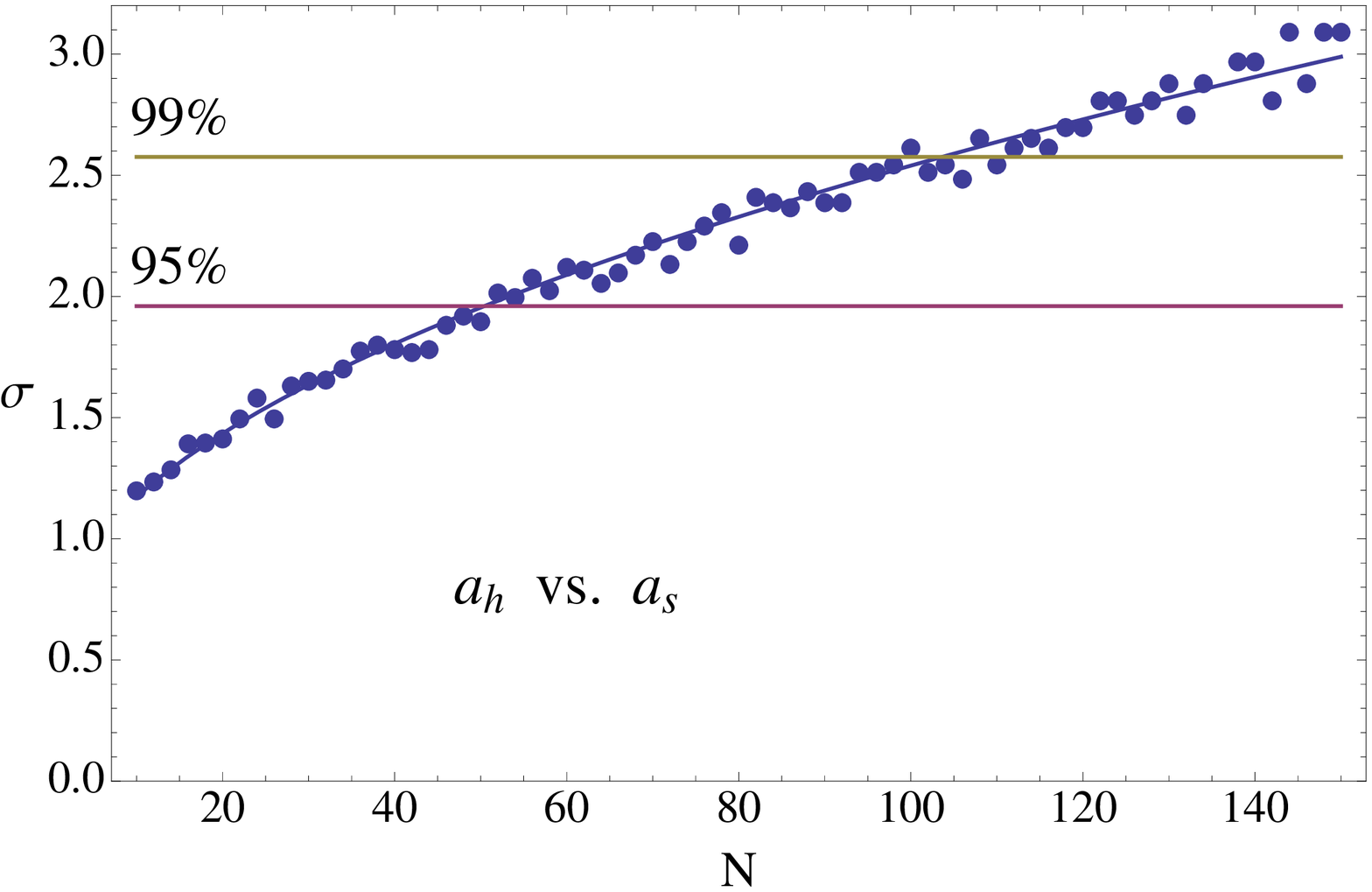}
\includegraphics[width=0.45\textwidth]{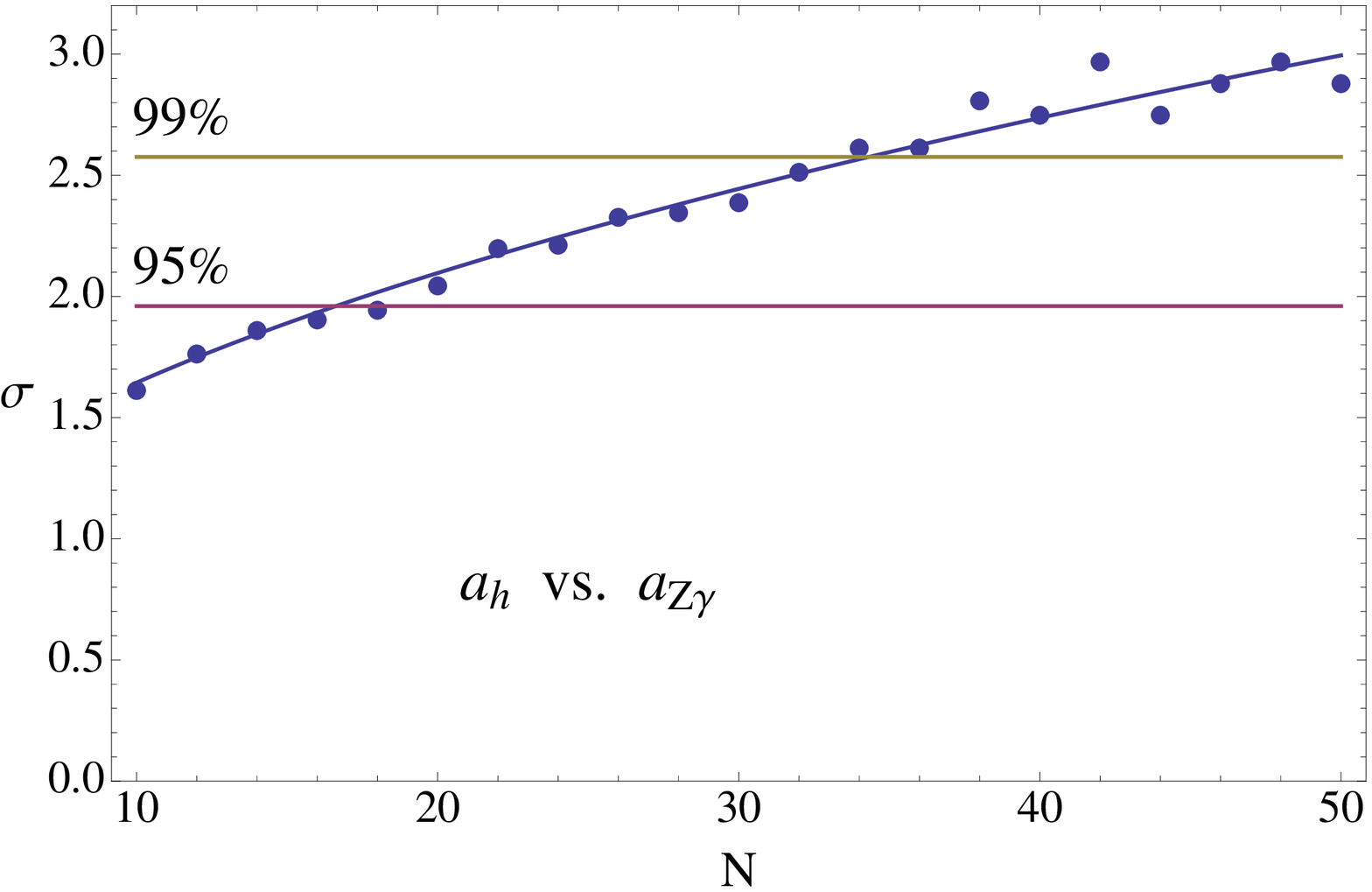}
\caption{Expected significance as a function of number of events in the case of $a_h$ vs $a_s$ on top, and  $a_h$ vs $a_{Z\gamma}$ on bottom. We use a different horizontal scale for the top and bottom plots because far fewer events are needed to discriminate $a_h$ from $a_{Z\gamma}$ than from $a_s$. We also fit with a function proportional to $\sqrt{N}$, which is the expected scaling. We mark the $\sigma$ value of 95\% and 99\% confidence level exclusion.  }
\label{fig:nev}
\end{figure}

This procedure is repeated many times for a range of numbers of events $N$ to obtain a significance as a function of $N$ for each hypothesis. We show this for the case where $a_1 = a_h$ and $a_2 = a_s$ or $a_2 = a_{Z\gamma}$ in Fig.~\ref{fig:nev}. We see that with ${\mathcal O}(50)$ events, we can distinguish renormalizable from nonrenormalizable coupling to $ZZ$ at 95\% confidence, and with ${\mathcal O}(100)$ events we can get a 99\% exclusion. The operator $a_{Z\gamma}$ can be distinguished from $a_h$ at 95\% confidence with as few as 20 events. The third possibility, which we do not show, is even easier; $a_s$ and $a_{Z\gamma}$ can be distinguished from one another at 95\% with just 10 events. 

\section{Conclusions and Outlook}
\label{sec:conc}

Testing the properties of the newly discovered resonance near 125 GeV is of utmost importance. While the rate and branching ratio data are consistent with the new particle being the Standard Model Higgs, direct tests of its properties are still essential. In this paper we have examined the discriminating power of events where the new particle decays to four leptons. These events can be used to measure the Lorentz transformation properties of this particle, but even if it is confirmed to be a parity even scalar, it still need not be the Higgs; it could couple to the gauge bosons via higher dimensional operators rather than via the renormalizable operator in the Standard Model. 

We have analyzed how well kinematic distributions in four lepton events can distinguish between different tensor structures of the coupling to gauge bosons. In particular, we looked a coupling directly to $Z_\mu Z^\mu$, as well as couplings to a pair of field strength tensors of the $Z$, and a coupling to the field strength of the $Z$ and of the photon. All three scenarios will produce one lepton pair near the $Z$ pole, while the other pair will have much lower invariant mass. We find that with ${\mathcal O}(50)$ signal events, a Higgs-like state can be discriminated from $ZZ$ field strength tensor couplings with 95\% confidence, while only 20 events are needed to make the same determination for field strength coupling to $Z\gamma$. This shows that the 2012 LHC run has excellent prospects to constrain the tensor structure of the new state's coupling to gauge bosons. 

While the four lepton final state is one of the most powerful for discriminating different scenarios, it would be interesting to look at kinematic variables in other final states. For example, in the decay to $WW^*$ where both $W$'s decay leptonically, the angles between the leptons and the transverse angles with missing energy will provide discriminating power, though this channel is difficult because of the large background. A search for decay to $Z\gamma$ where the photon is on-shell would give a direct measurement of the $a_{Z\gamma}$ coupling to the $Z\gamma$ field strength operator given in Eq.~(\ref{eq:zgam}). If this mode is in fact observed, kinematic analysis of final states in that channel could further uncover the nature of the new particle. 

There are many ways to both directly and indirectly learn about the couplings of the new state. We have argued that even if the state is a parity even scalar, it could be decaying to four leptons in a very non-Higgs-like way, possibly even through $Z\gamma^*$ instead of $ZZ^*$. Therefore, we hope that this work will inspire new measurements which can have strong discriminating power in the very near future. 

\vskip 0.3 cm

\noindent \textbf{Note Added:} While completing this paper, we received Refs.~\cite{Bolognesi:2012mm,Boughezal:2012tz} which discuss similar ideas.  

\vskip 0.3 cm

\noindent
We thank  Zackaria Chacko, Kyle Cranmer, Andrei Gritsan, Ian Low, and Kirill Melnikov for useful discussions. We also thank Markus Schulze and Nhan V. Tran for their help with the Johns Hopkins Monte Carlo, and Kunal Kumar for help with Madgraph. DS is supported in part by the NSF under grant PHY-0910467 and gratefully acknowledges support from the Maryland Center for Fundamental Physics. DS is also supported in part by the Project of Knowledge Innovation Program (PKIP) of Chinese Academy of Sciences, Grant No. KJCX2.YW.W10.

\end{document}